\documentstyle[aaspp4,12pt]{article}
\received{}
\accepted{}
\def\REF{\par\noindent\hangindent 20pt}

\def\rg{R$_{\rm g}$\/}
\def\ltsima{$\; \buildrel < \over \sim \;$}
\def\simlt{\lower.5ex\hbox{\ltsima}}            % < over MMM
\def\gtsima{$\; \buildrel > \over \sim \;$}

\def\simgt{\lower.5ex\hbox{\gtsima}}            % > over MMM

\def\hiha{{\sc Hi~H}$\alpha$\/}

\def\feka{{\sc Fe~K}$\alpha$\/}

\def\cm3{cm$^{-3}$\/}

\def\o4363{{\sc{[Oiii]}}$\lambda$4363\/}

\def\kms{km~s$^{-1}$}
\begin{document}
\title{\sc Disk Models for MCG-06-30-15: The Variability Challenge}
\author{J. W. Sulentic\altaffilmark{1}, P. Marziani\altaffilmark{2}, and M.
Calvani\altaffilmark{2}}

\altaffiltext{1}{Department of  Physics and Astronomy, University of
Alabama, Tuscaloosa, USA;
e-mail: giacomo@merlot.astr.ua.edu}

\altaffiltext{2}{Osservatorio Astronomico di Padova, vicolo
dell'Osservatorio 5, Padova, Italy}

\begin{abstract}
Recent observations have shown that the \feka\ line profile of the
Seyfert~1 galaxy MCG-06-30-15 is strongly variable.  We attempt
accretion disk model fits to the  \feka\ line profiles in a high, low
and medium continuum luminosity phase of this source. During the
monitoring by Iwasawa et al. (1996) a broad red-shifted component
remained reasonably constant while a narrower component  at $\approx$
6.4 keV strongly responded to continuum changes. Physically consistent
fits are possible if the  index $\xi$\ of the power-law  emissivity
changes from 0.7 (high phase) to 3.0 (low phase).

The shape of the red-shifted component at low phase is crucial to the
disk model interpretation. We suggest that the actual shape may be a broad redshifted Gaussian.
Three lines of evidence support the
interpretation of the \feka\ line as multicomponent, beyond the lack of
correlation in the response to continuum changes of the red and blue
components in MCG -06-30-15. (1) We show that the strong concentration
of narrow peak centroids at 6.4 keV is inconsistent with expectations of
a random distribution of disk orientations. (2) The average \feka\ profile
for a sample of 16 mostly Seyfert~1's suggests a natural decomposition
into two Gaussians one unshifted/narrow and the other redshifted/broad.
(3) Evidence for emission in excess to the expectation of disk models on
the high energy side of the \feka\ profile is both a challenge for low
inclination disk models and support for the two component decomposition.
\end{abstract}


\section{Introduction}

In an earlier paper (Sulentic et al 1997; hereafter S97)  we considered  the
problems associated with models that see the  X-ray \feka\ line emission at
$\approx$ 6.4 keV in Seyfert~1 galaxies as arising from fluorescence  reflection
(or emission) from an accretion disk.  A broad and redshifted \feka\ emission
feature is  observed in the spectra of many Seyfert~1 galaxies. The situation 
for other AGN classes is less well defined  and will not be  considered here.
Seyfert~1 emission profiles with a  narrow unshifted peak  at rest energy
$\approx$ 6.4 keV,   and a broad red-shifted wing extending down to $\approx$
4.5 keV are most suggestive of an accretion disk line profile (see e.g. Tanaka
et al. 1995). A great deal  of effort has been expended towards observing them
and fitting their spectra with disk models (see S97 for references). 
%Most of
%the pre-S97 fits were unconstrained in the sense that critical parameters
%such as inner radius,  inclination and emissivity law were simply chosen
%to minimize the $\chi^2$\ of the fit. In this way virtually all Seyfert~1
%sources with a detected line  have been fit with disk models.

In S97 we attempted to build a disk illumination model that could simultaneously
produce both \feka\ and optical Balmer lines. The motivation was to somehow
constrain the wide dispersion in disk model fits that have been published for
Seyfert~1 galaxies. 
%Our model has the
%advantage that disk inclination is the only  free parameter; all other
%parameters, including the emissivity law, come from the model.
While we could reproduce the profile widths and approximate emitting radii
consistent with observations, our model fits to the observed Balmer line
profiles were especially poor. In addition there were cases of severe 
disagreement between inclinations derived for \hiha\ and \feka. The diversity in
Seyfert~1 line profiles does not allow us to obtain a convergence in disk model
parameter space. Nandra et al. (1997a,b) also found empirically that no single
emissivity law could account for the diversity in line profiles among their
sample of 16 mostly Seyfert~1's. These problems may be more easily overcome if
one is prepared to consider a significant flux contribution from a second
non--disk source. 

Recently a challenge to disk models has arisen from the most studied Seyfert~1
galaxy MCG-06-30-15 (Iwasawa et al. 1996; hereafter I96). The \feka\ line
profile appears to change dramatically in response to continuum variations. In
\S \ref{data} we summarize the variability data and its challenge to the line
emitting accretion disk interpretation for \feka\ posed by MCG-06-30-15. We
attempt a  solution to  all variability phases in \S \ref{expla}.   In \S
\ref{compo} we consider the evidence for  the composite nature of the \feka\
profile,  and  in \S \ref{basta} we briefly discuss the implications for  disk
models. 

\begin{figure} 
\figurenum{1}
\plotone{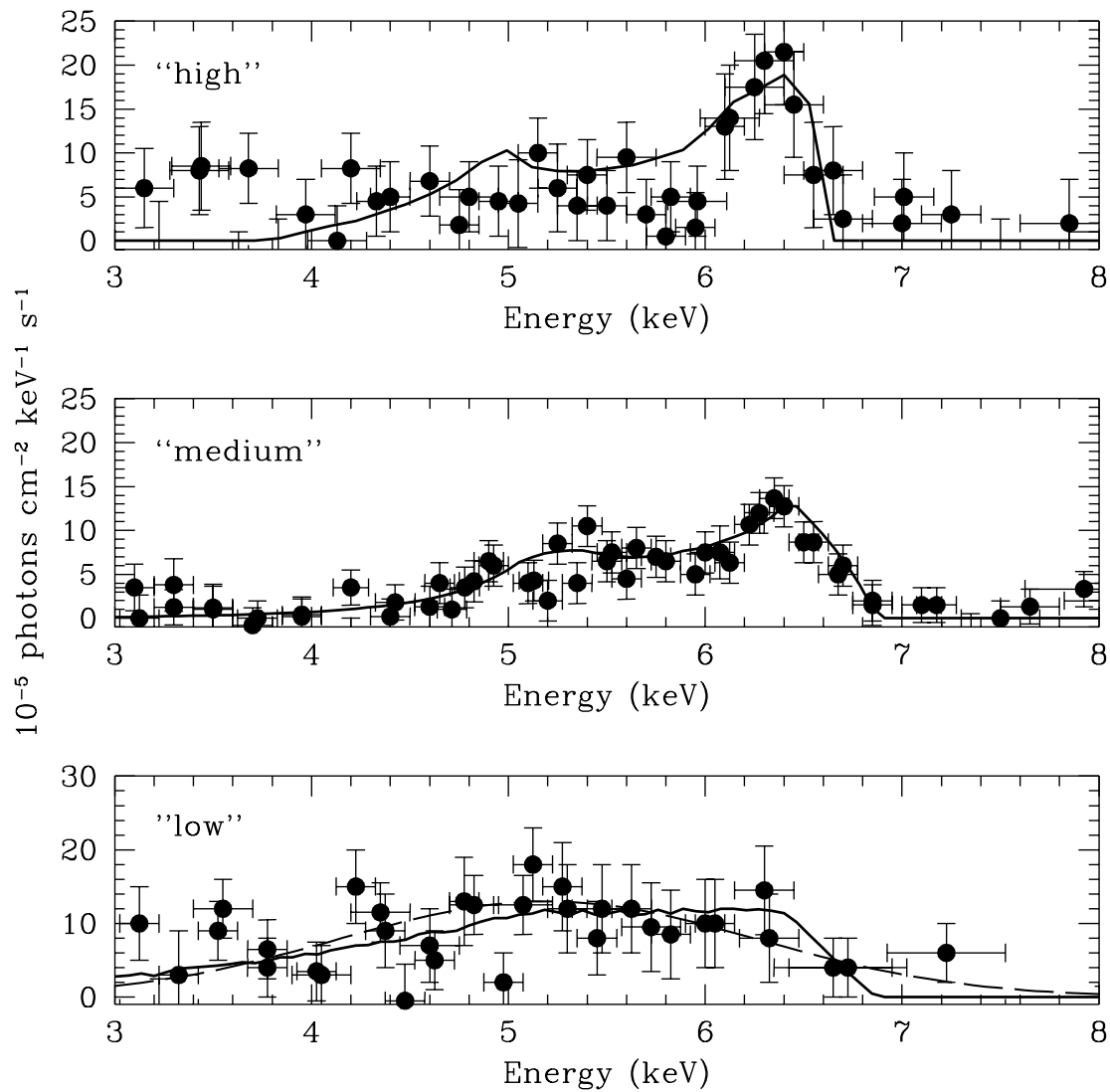}
\caption[1]{
\feka\ line profiles (filled circles) observed by Iwasawa et al. and best
fitting disk model profiles (filled line). See text and Table 1 for model
parameters. Ascissa units are rest energy in keV; ordinate units are 10$^{-5}$
photons cm$^{-2}$ keV$^{-1}$ s$^{-1}$.  Also shown (dashed line) is a gaussian
profile.}\end{figure} 

\section{Interpretation of Variability Data for MCG-06-30-15
\label{data}}

I96 report extensive observations of MCG-06-30-15. This source underwent
considerable variations in line and continuum intensity during a period of a few
days. I96 divide the data  and generate \feka\ profiles for high, medium and low
continuum phases (see Figure 1). We consider each of the phases beginning with
the medium one when the profile showed its most typical structure. In that 
phase the line is very similar to the \feka\ profiles for:  (1) MCG-06-30-15 as
shown by Tanaka et al. 1995) and (2) the average of 16 sources (mostly
Seyfert~1's) presented by Nandra et al (1997b). 

\begin{enumerate}
\item Medium phase: The line profile shows a strong and narrow
(unresolved at ASCA resolution)  peak centered at 6.4 keV along with a
broad ($\sigma$= 0.6 keV)  red wing extending down to $\sim$ 4keV. The
EW of the broad wing ($\sim$ 200-400eV) is approximately double that of
the unresolved peak in this phase.
%A structure within the red wing at ... is assumed due to instrumental
%and/or reduction artifact.
Tanaka et al. (1995) found a Kerr metric solution for the profile with 
accretion disk inner radius R$\rm
_{in}$=4.7R$\rm _g$ (R$\rm _g$ = GM/c$^2$), 
inclination i= 27$^\circ$ and $\xi$= 4.5 (where
the emissivity is described by a power-law $\rm \epsilon(r)\propto r^\xi$). 
I96 derive 7.6, 30$^\circ$  and 3.0 respectively. 

\item High phase: The narrow blue peak is considerably stronger when the
continuum is at highest level. In principal the red wing becomes weaker or,
even, disappears. The confidence contours around the best solution are
marginally consistent with both zero and unchanged flux. I96
fitted a disk model to the narrow peak which requires a very flat emissivity law
$\xi$= 1 and emission out to 1000 R$\rm _g$. This fit would be physically
inconsistent with the adopted medium phase fits. A solution more consistent with
the medium (and low phase) fits can be found if we assume that the red wing is
still present. 

\item Low Phase: The narrow peak disappears when the continuum is at lowest
level. There is either a weak residual peak blended with the broad red feature
or else the broad redshifted component is all that remains. I96
found a Kerr metric solution for this phase with R$\rm _{in}$= 1.24 R$\rm _g$
and  $\xi\sim$2.7. The shape of this low phase profile is not well defined.
Taken at face value it is reasonably flat over the 4-6.5 keV interval with a red
shoulder extending down to 3 keV or below. The low phase is very poorly sampled
blueward of 6.5 keV. 
\end{enumerate}

The variations therefore suggest: (a) a positive correlation between
continuum and blue peak
changes as well as (b) marginal evidence for a red wing-continuum
anticorrelation.

\section{Modeling the Changing \feka\ profile in MCG -06-30-15
\label{expla}}

An important aspect of the model fits to MCG-06-30-15 is that they apparently
require Kerr metric solutions while most previous fits used a Schwarzschild
metric (they are indistinguishable beyond 20 R$\rm _g$). In other words the best
data are driving us towards an extreme disk solution. Two challenges immediately
arise for models fits to  MCG-06-30-15: (1) producing a model that accounts for
all three phases with derived parameters that are physically
consistent/plausible and (2) accounting for the  lack of correlation between
blue and red component during continuum changes. 

Actually one must deal with two possible interpretations of the bright phase, a
strong blue peak with and without a weak red wing. In other words the most
unambiguous correlation is a positive one between blue peak and continuum flux.
The main confusion connected with the low phase involves the real shape of the
redshifted profile. We consider that in the next section.    A possible third
challenge involves fitting the smooth high energy  wing of the profile with a
disk model, but this also relates to discussion in the next section. Most models
produce a very sharp drop on the high energy side, a product of the effects of
gravitational redshift and Doppler boosting. We considered the  evidence for a
high energy smooth  wing in S97. 

In S97, we computed the  \hiha\ and \feka\ line profiles  produced by an
illuminated accretion disk (\feka\ and \hiha). The basic assumption of the model
is that a halo of free electrons scatters part of the continuum toward the disk.
We computed the scattered flux without taking into account relativistic effects
(like returning radiation). Apart from this approximation, and for
approximations on the vertical structure of the disk, the correct relativistic
treatment was used for all other computations in the model: the radial structure
of the disk was appropriate for the Kerr metric with 
specific angular momentum  a/M =
0.998 (c=G=1). The resulting cold  \feka\ surface emissivity 
 (a power-law  with radial emissivity index $\xi = 1.8$) was then
used to compute the line profile. We included all effects of relevance in the
Kerr metric, using the code developed by Fanton et al. (1997) to compute the
\feka\ line profile. 

In this paper we  attempt to reproduce the different \feka\ profiles
corresponding to different continuum luminosities. I96 isolated three
different profiles, typical of  ``low'', ``medium'', and ``high'' continuum
luminosity. In Figure 1 we show the I96 data  with superimposed model
profiles. We assume i=30$^\circ$.
Other model and  disk fit
parameters are in Table 1. The fits to all three luminosity phases are statistically
satisfactory, with normalized $\chi^2_\nu \approx 1$.  For each luminosity phase (listed in the  first
column) we report $\chi^2$, number of degrees of freedom, $\chi^2_\nu$, radial
emissivity index $\xi$,  disk inner and outer radii R$\rm _{in}$, R$\rm _{out}$ 
%(both in units of R$_g$).  
Dabrowski et al. (1997) obtained a
very similar emissivity law (with $\xi = 3.5$) inverting the
line profile to obtain the disk emissivity profile.

If a hot halo indeed scatters radiation toward  a cold disk, the change of index
in the radial emissivity law can be  understood at least in qualitative terms.
The emissivity change reflects a change in the distribution of the scattering
matter: when the continuum is high, radiation pressure pushes the scattering
matter outward; when the continuum luminosity is at a minimum, the scattering
halo may not be fully supported against the massive black hole gravity, and the
radial distribution of matter in the halo may follow a steeper  power-law. 

The success of our model requires a significant red component even during  the
high variability phase. Fits to an isolated blue  peak during this phase are
physically incompatible with intermediate and low phase ones. Our fit to the red
part of the high phase profile is 30-50 \%\ too high. This is reflected in the
larger $\chi^2$. It is unclear how much of this discrepancy might be related to
uncertainties in the continuum fit. The observational results are consistent
with  a rapid response to the continuum by the blue peak plus a weak (or zero as
favored by our model) anticorrelation by the red  wing. This interpretation is
apparently strongly driven by the response of the blue peak to large continuum
fluctuations. When the largest flare and lowest minimum are excluded from the
dataset I96 find almost the opposite result. The broad red
wing correlates  with more modest/frequent fluctuations  in the continuum while
the data are consistent with a constant blue peak. In other words the line
response may depend on the amplitude of the continuum fluctuation.  The  data
without the strongest continuum fluctuations included would require a model with
two distinct emitting components because disk models could not account for a
constant blue (Doppler boosted) component. We consider the observational
evidence for a two component profile in the next section.

\begin{deluxetable}{ccccccc}
\tablewidth{27pc}
\tablecaption{Disk Model Fits to Fe K$\alpha$\ Line Profile of MCG-6-30-15}
\tablehead{
\colhead{Phase} & \colhead{$\chi^2$} & \colhead{D. of F.} & \colhead{
$\chi^2_\nu$} &
\colhead{$\xi$} & \colhead{R$\rm_{in}$/R$\rm _g$ } &\colhead{R$\rm_{out}
$/R$\rm _g$}}
\startdata
high    & 54.6  &  40  &  1.37  &  1.5  &  6.00  &  15.0 \\
medium  & 63.6  &  51  &  1.25  &  0.7  &  1.23  &  20.0 \\
low     & 35.3  &  31  &  1.14  &  3    &  1.23  &  20.0 \\
%\tablecomments{}
%\tablerefs{}
\enddata
\end{deluxetable}
\section{ A Composite \feka\ Profile: the Variability
Challenge\label{compo}}

Our model fits to MCG-06-30-15 have followed the precedent set by  other workers
by modeling the entire profile. The underlying assumption is that  most or all
of the line flux arises from a single-component  \feka\ line(6.4, 6.7 or  6.9
keV) produced in a single emitting  region (the accretion disk). In S97 we
considered an alternate model for  the production of the  entire profile 
(infalling clouds in a bi-conical geometry).

The exact shape of the low phase profile in MCG-06-30-15 is crucial to the
success of the disk models. The assumption of a single emitting region  is also
fundamental. As it stands the low phase line  profile could be fit by a wide
variety of shapes. The S/N is low and  and the high energy wing is undersampled.
We propose that the actual shape may be Gaussian and that the red peak is
independent of the blue one. Recently, several authors have followed an empirical approach  to the \feka\ line profile fitting, employing single or multi-Gaussian profiles  (i.e., Grandi et al. 1997; Weaver et al. 1997).
There are several lines of evidence to support
this interpretation for MCG -06-30-15. The first  is the lack of correlation between the red and
blue component responses to  continuum changes. Our model discussion above
suggests that this is not an insurmountable problem if our assumptions about the
high (red component present$\approx$ same level as medium phase), intermediate
and low  (negligible/absent blue peak, non-Gaussian red profile) phase data are 
correct. 

A more serious problem is illustrated in Figure 2 where we show the distribution
of blue peak centroid energy versus profile width (the blue peak is unresolved
with ASCA). We see a strong concentration of  centroid energies at 6.4 keV. Also
plotted are the predictions of disk models  similar to the ones implied by the
MCG-06-30-15 fits. We show the model fits for the range of inclinations
0-45$^\circ$ expected for Seyfert~1 galaxies in a standard unification scheme.
It is obvious that 6.4 keV has no special significance for the disk models; a
range of centroid energies is expected  if we are viewing line emitting
accretion disks at random inclinations.  An energy of 6.4keV has special
significance only as the rest wavelength of  fluorescence reflection \feka. The
only escape from this excess of 6.4 keV  peaks is to argue that the current
observational data are strongly biased  towards a narrow range of disk models
near 6.4 keV, for instance, because  more inclined profiles are broader and
hence less easy to detect at  high S/N. Similar results are obtained from a plot
of narrow peak energy versus  centroid (measured on the broad base). 

The remaining evidence brings us back to the exact shape of the low phase
profile (when the narrow peak is assumed absent). We suggest that the exact
shape can be inferred from the composite spectrum presented in  Nandra et al.
(1997). The shape of the average spectrum is robust to the  exclusion of
MCG-06-30-15. The simplest and most natural fit to the average  spectrum
involves two Gaussian components: (1) a narrow (FWHM$\simlt$0.25 keV) component
centered at \feka\ rest energy (E$\rm _n$= 6.4$\pm$0.05 keV) and (2) a broad
(FWHM = 1.6 keV) redshifted (E$\rm _b$ = 5.9$\pm$ 0.1 keV). 

Another piece of evidence that supports this interpretation and profile
decomposition involves a smooth wing on the \feka\ line high energy side. The 
wing appears to grow in significance as the S/N of the MCG profile increases.
Thus it is best seen in the high and intermediate phases when  the narrow peak
is visible. Is it the wing  an independent high-energy component  while the 
much broader red component arises in a disk?  The I96 data are ambiguous here,
one plot shows a high energy wing while  another shows a much steeper drop. Not
surprisingly the latter is used for  the Dabrowski et al. (1997) fits because
most Kerr models so far explored show  a steep drop on the high-energy side. In
the Nandra et al. (1997b) average spectrum,  the high-energy wing looks like a
natural extension of the broad red component  under the blue peak. In S97 we
showed that a photoionized disk can produce  enough hot \feka\ to account for a
blue wing. In that case the blend  of hot and cold \feka\ emission must conspire
to produce the observed  smooth average profile. 

Figure 1 shows a single Gaussian fit (dashed line) to the ``low'' phase profile.
The Gaussian parameters are quite similar to those found for this  component in
the average profile of Nandra et al (1997). We show that a  fit as good as  the
one obtained with the disk  model can be achieved, with $\chi_\nu \approx 1.08$
obtained for peak  energy $\approx 5.2$ keV and dispersion $\sigma \approx .75
$keV. If the redshifted component is (a) independent and (b) Gaussian in shape,
then there is no obvious disk model that can account for it. 

\begin{figure}
\figurenum{2}
\plotone{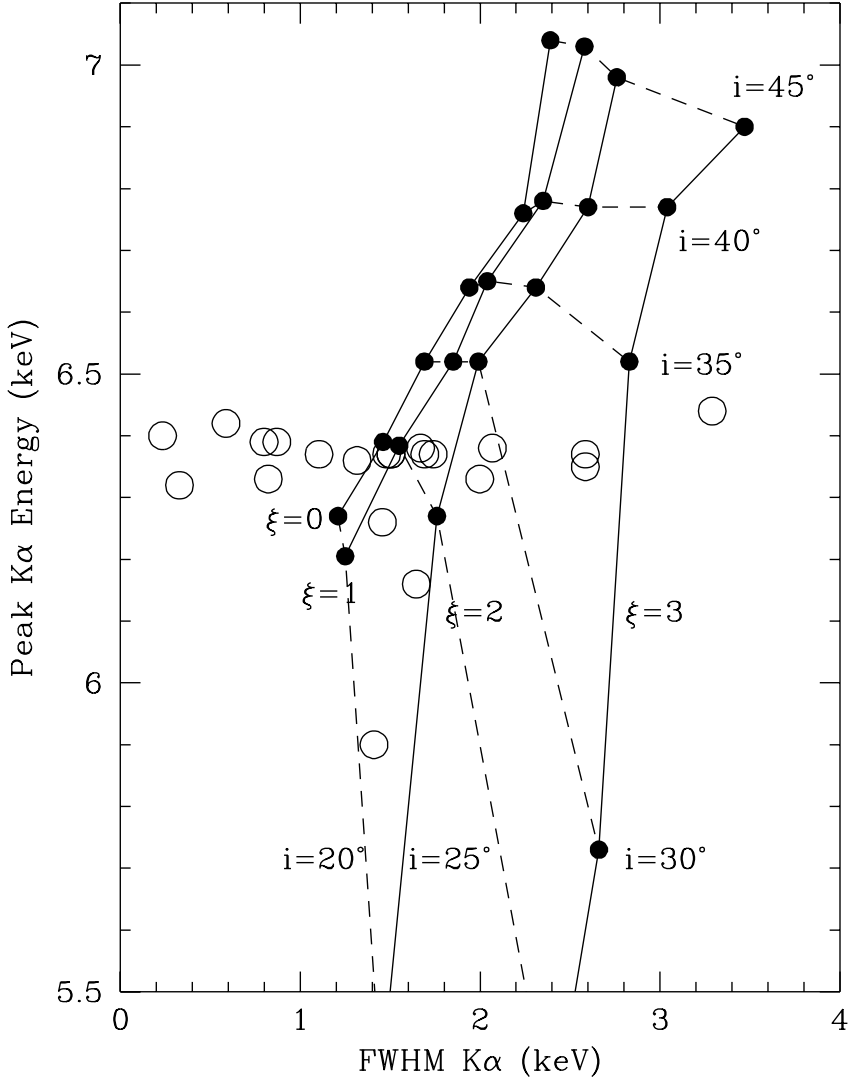}
\caption{
\feka\ narrow peak rest energy in keV versus line FWHM in keV. Open circles:
measurements  by Nandra et al. 1997. Model profiles (filled circles) were
computed  assuming a=0.9981; R$\rm _{in}$ =  1.233 R$\rm _g$, and R$\rm _{out}$
= 20.0 R$\rm _g$, with different emissivity power-law   index $\xi =
0.0,1.0,2.0,3.0$, and  for 6 different values of inclination (i=20$^\circ$,
25$^\circ$, 30$^\circ$, 35$^\circ$, 40$^\circ$, 45$^\circ$). Filled line joins
points with same $\xi$; dashed line joins points with same
inclination. In case the \feka\ profile is double-peaked, the maximum has been set at the height of the red peak, to ensure that the FWHM
 is a measurement of the width of the broad red-shifted feature. }\end{figure} 

\section{Conclusion \label{basta}}

Accretion disk model fitting to the variable \feka\ profile of MGC -06-30-15
require a Kerr black hole: even if only the low continuum phase \feka\ profile
cannot be fitted by disk models around a Schwarzschild black hole, the angular
momentum of a massive black hole obviously cannot change on timescales of
$\approx 10^5$ s. As the inner disk radius for a  Kerr black
hole 
with a/M = 0.998
is $\approx$ 1.23 \rg, and as the region where the ``Doppler boosted peak'' 
is formed
occurs at $\rm R \approx 5-20$ \rg, different illumination  of the disk
following changes in the continuum luminosity explains the strong change
observed in the narrow peak at 6.4 keV (Plate 1 of Fanton et al. 1997 shows that
it is not so for a Schwarzschild black hole). On the other hand, without
attempting a physical interpretation of the \feka\ line profile, the strong
variations of the narrow peak, along with the possibility of little or no
variation in the redshifted broader line part, hint at two independent
 components.
The two component hypothesis is reinforced by the detection of the narrow peak
at 6.4 keV in the wide majority of cases. This statistical difficulty for
accretion disk models must be understood, before  accretion disk models can be
accepted. 

The formulation of a physical model for a multicomponent \feka\ line goes beyond
the aims of the present letter. We can, however, speculate on several
possibilities. A possibility, recently revived by Misra \&\ Khembavi (1998), is
that  broadening could occur by  Compton scattering the photons of an
intrinsically narrower \feka\ line in a corona  of size  $\approx$ 300 GM/c$^2$,
much larger than the size of the region of continuum formation.   Another
possibility is that the narrow peak could be associated to the BLR: with an
intrinsic width of 30000 \kms\, \feka\ emission from the BLR should appear as an
unresolved peak. The maximum line shifts in the BLR are  too small to be
detected with ASCA, so that the peak would appear always at about 6.4 keV (as
expected for cold iron emission). Clearly the detection of strong changes in the
narrow peak with no apparent time delay (along with some difficulties raised by
S97) challenges these interpretations, but a longer time coverage would be
required to rule them out. The  redshifted broader component may be, on the other
hand, associated with the region of continuum production. Observations of a
cut-off in the X-ray continuum of Seyfert galaxies at energy $\approx$ 600 keV
suggests the presence of  a thermal  corona above the surface of the accretion
disk (e. g. Zdziarski et al. 1995; Haardt \&\ Maraschi 1991;  also S97). \feka\
emission from highly ionized iron in the corona could be shifted to the observed
rest energy by gravitational redshift. 

\section{Acknowledgements}

We are indebted to C. Fanton for allowing us to use his program. It was
invaluable for the success of these computations. JS acknowledges financial
support from the Italian CNR. 

\newpage
\section{References}
\REF
Dabrowski, Y., Fabian, A. C., Iwasawa, K., Lasenby, A. N., \&\ Reynolds,
C. S.,
1997, MNRAS, 288, L11
\REF
Fanton, C., Calvani, M., de Felice, F., \&\ Cadez, A., 1997, PASJ, 49, 159
\REF
Grandi, P.,  Sambruna R. M., Maraschi L., Matt G., Urry C. M., \&\  Mushotzsky R. F., 1997, ApJ, 487, 636
\REF
Haardt, F.,  \&\ Maraschi, L., 1993, ApJ 413, 507
\REF
Iwasawa, K.,  et al., 1996, MNRAS, 282, 1038 (I96)
\REF
Misra, R.,  \&\ Kembhavi, A. K., 1998, ApJ, in press.
\REF
Nandra, K., George, I. M., Mushotzsky, R. F., Turner, T. J., \&\ Yaqoob,
T.,
1997a, ApJ, 476, 70
\REF
Nandra, K., George, I. M., Mushotzsky, R. F., Turner, T. J., \&\ Yaqoob,
T.,
1997b, ApJ, 477, 602
\REF
Sulentic, J., Marziani, P., Zwitter, T., Calvani, M., \&\
Dultzin-Hacyan, D., 1997, ApJ, submitted (S97)
\REF
Tanaka, Y., et al., 1995, Nature, 375, 659
\REF
Weaver, K. A., Yaqoob, T., Mushotzky, R. F., Nousek J., Hayashi I., \&\ Koyama, K., 1997, ApJ 474, 675
\REF
Zdziarski, A. A., Johnson, W. N., Done, Chr., Smith, D., McNaron-Brown,
K., 1995, ApJ, 438, L63
\newpage

%\end{document}
\newpage

%changed

\end{document}